# Terahertz Frequency Comb High-Resolution Heterodyne Spectrometer

Francis Hindle, Alexandra Khabbaz, Anthony Roucou, Jean-Francois Lampin, Gaël Mouret

*Abstract*— We demonstrate the advantages of THz frequency combs for high-resolution spectroscopy. This benefits from wide spectral coverage and the exact knowledge of the frequency position of each comb component. Heterodyne detection combined with a fast Fourier spectrometer enables rapid and simultaneous measurement of more than 80 frequency comb modes covering a 7.5 GHz bandwidth. A spectrum is obtained in under 20 minutes yielding a uniform resolution of 70 kHz. This new setup has been validated by recording more than 150 lines of methanol around 723 GHz, and represents a new solution to exploit THz frequency combs for high-resolution spectroscopy.

## I. INTRODUCTION

THE analysis of gases using mm-wave and THz radiation is of particular interest as the rotational lines can be very narrow providing a high degree of molecular discrimination. This has recently been demonstrated to measure the weak lines of $CF_4$, leading to an improved global fit and line listing [1]. Molecules with stronger line intensities have been quantified under trace conditions for industrial [2], food safety [3], or medical [4] applications. Such measurements are most easily made with an instrument that has a spectral resolution that is finer than the Doppler broadening of the molecular lines in question, for example $H_2S$ which has strong lines around 600 GHz and displays a Doppler broadening of 0.64 MHz (Half Width Half Maximum).

Frequency Domain Spectroscopy (FDS) and Time Domain Spectroscopy (TDS) are the two approaches routinely employed in the mm-wave/THz band. FDS uses a continuous wave spectrally fine source with a sensitive detector, the source is frequency tuned and the spectrum built point-by-point. Both frequency upconversion [5], [6] and downconversion [7] sources have successfully been used. The advantage FDS is that an excellent resolution can be easily obtained and the frequency positions can be referenced to a primary standard, the data acquisition process is however time consuming as each line is individually measured point by point.

THz-TDS uses a femtosecond laser to create a short wideband THz pulse using a rapid semiconductor photoswitch [8]. Once the pulse has propagated through that material to be measured it can be detected by combing it with a fraction of the optical pulse in a second semiconductor device. A delay stage used to control the arrival time of the optical pulse and allows the received THz field to be characterised as a function of time. The Fourier transform of the time domain data yields the spectrum. Here the advantage is the wideband spectral coverage of this so-called multiplex technique that simultaneously measures many spectral channels. Unfortunately, the frequency resolution available is limited by the delay stage via the available time window, values down to 500 MHz can be achieved. A super-resolution data treatment has been demonstrated to providing a resolution of 50 MHz for a sparse spectrum [9]. THz-TDS instruments are therefore not suitable for high-resolution spectroscopy and so cannot achieve the same degree of molecular discrimination compared to FDS.

The advent of frequency combs (FC) obtained by accurately stabilising the repetition rate of femtosecond lasers has led to many new possibilities for frequency metrology and applications [10], [11]. Indeed, in the case of the THz band a FC is generated from an optical FC using a photoswitch, the THz-FC generated is a harmonic comb which unlike its optical counterpart does not have an offset frequency. The spectral width of an individual mode of the THz-FC is sufficiently narrow that if correctly harnessed should be able to provide an excellent frequency resolution [12]. ASynchronous Optical Sampling (ASOPS) or Dual Comb Spectroscopy have been applied to THz spectroscopy using two FCs with a small difference in the repetition rates. This approach is similar to TDS but no mechanical delay stage is required and acquisitions with larger time windows can be undertaken enabling the individual THz-FC modes to be resolved. Although a single

The authors would like to thank Thales Research and Technology, the Université du Littoral Côte d'Opale and the Région Hauts-de-France for funding the doctoral studies of Alexandra Khabbaz. The authors would also like to acknowledge the financial support of the French Agence Nationale de la Recherche via TIGER (ANR-21-CE30-0048) and HEROES (ANR-16-CE30-0020). This work was carried out within the framework of the Contrat de Plan Etat-Region (CPER) WaveTech@HdF, which is supported by the Ministry of Higher Education and Research, the Hauts-de-France (HdF) Regional Council, the Lille European Metropolis (MEL), the Institute of Physics of the French National Centre for Scientific Research (CNRS) and the European Regional Development Fund (ERDF).

Francis Hindle, Alexandra Khabbaz, Anthony Roucou, and Gaël Mouret the authors are with the Université du Littoral Côte d'Opale, UR 4493, LPCA, Laboratoire de Physico-Chimie de l'Atmosphère, F-59140, Dunkerque, France. (e-mail: francis.hindle@univ-littoral.fr).

Alexandra Khabbaz, is with the Université du Littoral Côte d'Opale, UR 4493, LPCA, Laboratoire de Physico-Chimie de l'Atmosphère, F-59140, Dunkerque, France and the Institut d'Electronique, de Microélectronique et de Nanotechnologie, UMR CNRS 8520 Avenue Poincaré, B.P. 60069, 59652 Villeneuve d'Ascq, France.

Jean-Francois Lampin, is with the Institut d'Electronique, de Microélectronique et de Nanotechnologie, UMR CNRS 8520 Avenue Poincaré, B.P. 60069, 59652 Villeneuve d'Ascq, France.

mode has a linewidth certainly finer than 0.1 MHz the discrete spacing at intervals of the repetition rate effectively degrades the useable resolution [13]. This can be overcome by interleaving multiple spectra taken with different repetition rates, a frequency resolution of 2 MHz has been demonstrated [14]. The advantage here is that the entire band is measured simultaneously however the data accumulation to obtain high-resolution spectra is however particularly time-consuming and the resolution is not uniformly distributed over the entire band. The potential of THz-FC to produce high-resolution data over a band of frequencies remains to be unlocked.

One alternative approach is to employ a sensitive heterodyne detection scheme [15]. Such receivers are commercially available and can offer a bandwidth up to around 40 GHz, so can simultaneously monitor a large number of FC modes. If this process can be combined with the tuning of the repetition rate, high-resolution data should be obtained with a continuous uniform coverage over the band in question. The difficulty here is the measurement and processing of the intermediate frequency (IF) that contains multiple FC modes over a large band. A heterodyne scheme has been demonstrated in the millimetre range with a single mode being monitored [16] with a superlattice mixer and a standard spectrum analyser. In this case the laser repetition frequency, $f_{rep}$, and the local oscillator (LO) frequency were simultaneously adjusted so that the FC mode was maintained at a constant IF. A standard electrical spectrum analyser (ESA) can therefore be used to continuously monitor the FC mode. A 100 MHz portion of the spectrum of $CF_3H$ at 289 GHz was obtained with a resolution of 100 kHz, requiring an accumulation period of 180 minutes. Here we have developed a multimode THz-FC spectrometer that operates in the band 500 – 750 GHz and is able to efficiently record a large number of THz-FC modes, producing high-resolution (~70 kHz) spectra covering 7.5 GHz in 20 minutes. This solution provides a frequency band almost 2 orders of magnitude larger in roughly one tenth of the time.

## II. Spectrometer

The THz-FC is generated by a femtosecond laser, however unlike a THz-TDS the discrete nature of the optical comb produced can be harnessed by the fine control of the repetition rate. A fibre laser (Menlosystems, C-Fiber) operating at 1560 nm, and equipped with a second harmonic generation unit to provide 780 nm was employed to provide an average power of 10 mW. The laser has a nominal repetition rate, $f_{rep}$, of 88 MHz and can be tuned over +/- 200 kHz, it is locked to a synthesiser that is referenced to a GPS timing signal (Spectracom EC20S). A Low Temperature Grown-GaAs dipole antenna converts the optical femtosecond pulses to a harmonic THz-FC, which in contrast to an optical frequency comb has no offset frequency.

Once the THz-FC has interacted with the gas sample a heterodyne scheme is used to isolate a spectral window centred at twice the LO frequency $f_{LO}$, Fig. 1. The heterodyne receiver is a sub-harmonic mixer (SHM) for the band 500-750 GHz, it uses the second sub-harmonic (m = 2), and has a typical insertion loss of 9 dB. It is fed by an amplified multiplier chain (Virginia Diodes, SGX) operating in the 250-375 GHz band and providing a frequency of $f_{LO}$. A second synthesiser drives the LO chain and is also referenced to the GPS timing signal.

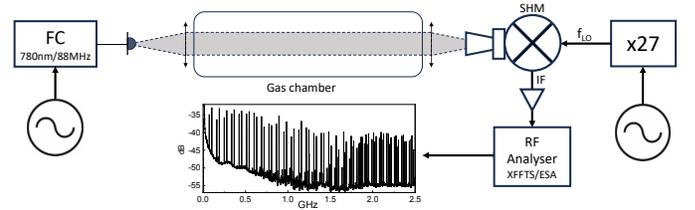

Fig 1. THz-FC heterodyne spectrometer. The THz-FC is generated from a stabilised femtosecond laser (Menlosystems, C-Fiber) focused on a LTG-GaAs dipole antenna. The THz beam is propagated through the gas cell (L=50 cm) and coupled to a sub-harmonic mixer (Virginia Diodes, WR1.5SHM) which is able to operate in the 500-750 GHz band. The SHM is pumped by a frequency multiplier driven by a microwave synthesiser (Virginia Diodes, SGX). The IF signal can be analysed either by the XFFTS or a conventional ESA. The example data is shown for an XFFTS. Both synthesisers are referenced to the same GPS timing signal. TPX lenses with 50 mm focal lengths were used to collimate the THz beam at the gas chamber entrance and to couple the beam to the mixer. The IF signal is amplified by two amplifiers (Minicircuits, ZX60-V62+).

The IF signal contains multiple beatnotes of the THz-FC mixing with the LO. If an ESA is used to analyse the IF signal the frequency centre, span and resolution are configured. The ESA sweeps across the span, the discrete nature of the comb implies that the signal is only accumulated at the THz-FC mode positions. To improve the accumulation efficiency the IF can alternatively be analysed by a Fast Fourier Transform Spectrometer (FFTS). The THz-FC is able to provide high-resolution information as each mode is spectrally narrow, the subsequent modes are however separated by $f_{rep}$. In order to obtain a uniform coverage at high resolution the data from a range of values of $f_{rep}$ should be combined. If the repetition frequency is scanned from $f_{rep}$ to $f_{rep} + \delta$ then all the THz frequencies at $\nu_{THz}$ can be covered as given in (1).

$$\delta \approx \frac{f_{rep}^2}{\nu_{THz}}$$

(1)

## III. Results and Discussion

*Single frequency comb mode*

The initial validation was performed using a standard ESA to measure a single THz-FC mode as it was progressively scanned across a molecular transition of $D_2O$ located at 722669.85 MHz. Small frequency steps can be applied to $f_{rep}$ without losing the lock between the laser and reference oscillator. An ESA spectrum was recorded for each value $f_{rep}$ as it was stepped from 88.0550 to 88.0556 MHz. This corresponds to real frequency range of 722.6674 to 722.6723 GHz for the THz-FC mode number 8207. Eight complete ESA spectra are shown as an example in fig. 2 for a gas pressure of 20 μbar. At this pressure a step size of $f_{rep}$ 6.25 Hz was employed achieving a sample spacing of 51 kHz. The ESA was configured with a resolution bandwidth of 50 kHz to match this value and minimise the acquisition time, a total of 96 ESA spectra are required to cover the 4.9 MHz. Decreasing the resolution bandwidth significantly slowed down the ESA measurement but indicated THz-FC mode had a FWHM finer than 1 kHz well suited for high resolution spectroscopy. Each of the acquired ESA spectrum was processed by straightforwardly extracting the maximum value. The data compiled in this way for a pressure of 20 µbar are presented in fig.2 along with a portion of the baseline taken with an empty cell. A Voigt absorption profile was fitted to the measured data giving a centre frequency of 722.66983 GHz +/- 10 kHz, this value is within 20 kHz of the tabulated value for this molecule which has previously been measured with an experimental uncertainty of 200 kHz [17], [18].

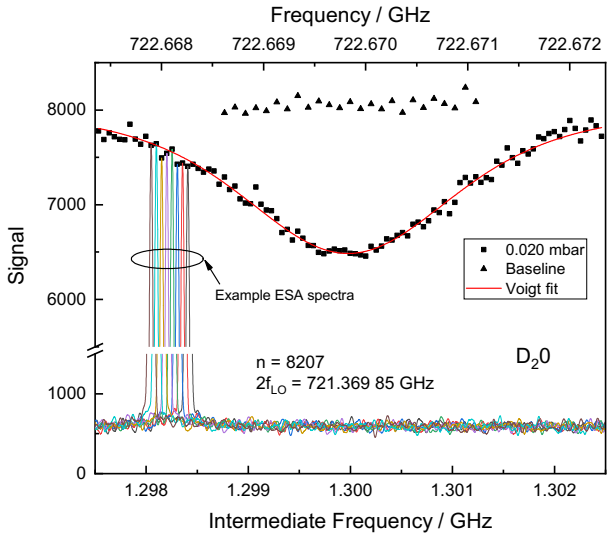

Fig 2. Spectrum of $D_2O$ ( $J_{Ka,Kc}$ = $5_{2,3}$ ← $4_{3,2}$ ) measured using mode 8207 of the THz-FC. Eight unprocessed ESA spectra (fine lines) are shown demonstrating the frequency steps of the FC-THz and the extraction of the maximum values to construct the complete spectra. Squares, data for 20 µbar with a $f_{rep}$ step of 6.25 Hz giving a total of 96 data points, this data has been fitted using a Voigt absorption profile (thick red line). The triangles show the baseline taken with an empty cell.

*Multiple frequency comb modes*

This initial data acquisition clearly demonstrates the full potential of this new setup. However, such an approach quickly becomes cumbersome and time consuming as the required spectral coverage is increased. An alternative is to use a FFTS that is sensitive wide-band digital frequency domain analysers that have been developed for radio astronomical applications. They are based on high-speed analogue to digital converters and field programmable gate arrays. The eXtended bandwidth FFTS (XFFTS) used provides 2.5 GHz of analysis bandwidth which is divided into 32768 equal frequency channels [19]. The advantage of the XFFTS, compared to an ESA, is that the data are accumulated on all the channels simultaneously. The example spectrum shown in Fig. 1 was obtained for a measurement duration of 1 second, the THz-FC modes are clearly visible with a contrast in excess of 10 dB compared to the noise floor.

The XFFTS used had a channel spacing of 76 kHz, if frequency steps of this size are applied to the THz-FC a spectrum can be constructed. A total of 1160 such steps are required to cover 88 MHz per mode, enabling complete and uniform coverage of the available IF analysis bandwidth. The XFFTS has a sampling frequency $f_s = 5\ GHz$, so its nominal bandwidth is 2.5 GHz, corresponding to the first Nyquist zone (DC to 2.5 GHz). Beatnotes at higher and lower frequencies than $2f_{LO}$ are detected to provide an initial THz coverage of 5 GHz. In addition, a number modes falling in the second Nyquist band of the XFFTS (2.5 to 5 GHz) are also detected and identified. They can be observed in the example in Fig. 1 at the right-hand side of the frequency scale as a higher density of modes. The response of these modes starts to decrease above 3.5 GHz (or below 1.5 GHz on the XFFTS frequency scale). The information contained in these modes allows the THz spectral coverage to be extended to around 7.5 GHz. With an accumulation time of 1 second for each value of $f_{rep}$, the complete spectrum can be acquired in a little under 20 minutes.

A large number of modes are easily identified as demonstrated by an example spectrum in fig. 3. The THz-FC mode closest to $2f_{LO}$ is mode number 8213 at an IF of 16.1 MHz. In this case the mode frequency is lower than $2f_{LO}$. A series of modes can be identified by decrementing the mode number, the lines are observed at intervals of $f_{rep}$. This series contains the modes from 8213 to 8185 before reaching the nominal IF frequency limit of the XFFTS. A second series incrementing the mode number from 8214 (at 72.0 MHz) to 8241 (at 2449.1 MHz) is observed for frequencies higher than $2f_{LO}$. Hence a total of 57 modes are identified in the nominal baseband (DC to 2.5 GHz) of the XFFTS.

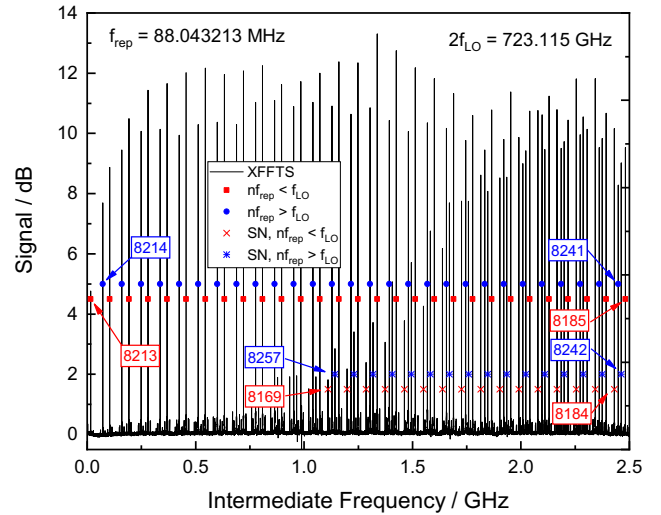

Fig 3. Example XFFTS spectrum after baseline removal by subtraction of spectrum measured without THz-FC. A total of 89 THz-FC modes are identified by the symbols. Red symbols for modes at frequencies lower than $2f_{LO}$, blue for those at higher frequency. The XFFTS baseband lines are shown by the filled symbols. The crosses indicate the modes in the second Nyquist (SN) band of the XFFTS. A selection of mode numbers are given at the start and end of the series.

*Data processing*

The recorded XFFTS data for a complete scan consists of 1160 spectra each containing 32768 values (total of 38 x 10$^6$ data points). Only the data at the THz-FC modes position contain any information, a data processing algorithm is therefore required and has two principal roles. Firstly, in a given spectrum it extracts the signal intensity at each of the THz-FC mode positions corresponding to, the value of $f_{rep}$ and the LO frequency, as given by (2). Secondly, it attributes the frequency of each mode using $f_{rep}$, the FC-THz mode number $n$, and $m = 2$ for the heterodyne receiver. All the XFFTS spectra are processed to produce a single file containing the measured intensity as a function of the frequency.

$$f_{IF} = |nf_{rep} - mf_{LO}| \quad (2)$$

A number of additional lines are clearly observed at frequencies from 1.5 to 2.5 GHz on the baseband frequency scale in fig. 3. These lines correspond to frequencies in the second Nyquist zone of the XFFTS, that is from 2.5 to 5.0 GHz. The frequency scale for these lines will run from 5.0 to 2.5 GHz instead of from 0.0 to 2.5 GHz for the nominal baseband scale. For example, the THz-FC mode number 8184 would have a standard IF of 2569.4 MHz which is above the 2500 MHz limit. As it is in the second Nyquist zone it will appear at 2430.6 MHz on the nominal frequency scale, as calculated by equation (3).

$$f_{IF} = f_s - |nf_{rep} - mf_{LO}| \quad (3)$$

This series can be continued until around 8169 where the amplitude becomes weaker and starts to approach the noise floor. Also in the second Nyquist zone is the series from 8242 to 8257, as indicated in fig. 3. A total of 32 additional modes are identified from the second Nyquist zone.

*Methanol spectrum*

A series of 1160 XFFTS spectra where recorded while steps of 9.28 Hz were applied to $f_{rep}$. The step size was chosen to match the XFFTS channel spacing at the targeted frequency of 723 GHz. The spectra measured for 50 μbar of methanol and an empty cell were processed, fig. 4. Here the signal strength is the contrast between the noise floor of the XFFTS and the THz-FC modes. If 85 modes are used then a coverage of 7.48 GHz is achieved with 98.6 x 10$^3$ data points.

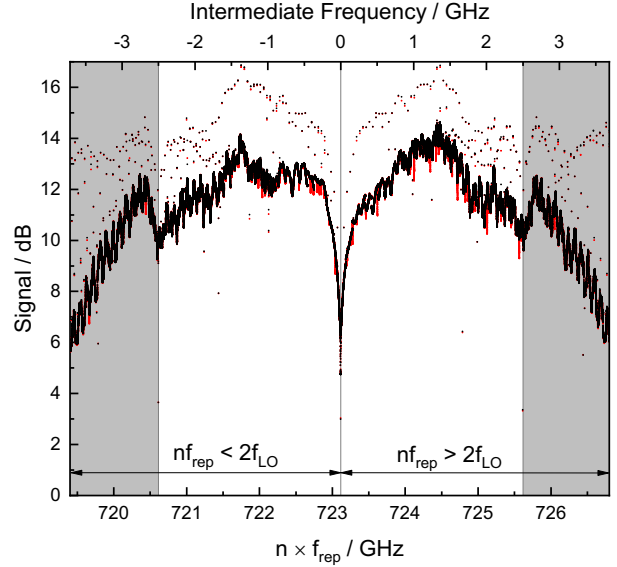

Fig. 4 Signal strength of the FC-THz modes with $2f_{LO} = 723.115\ GHz$ indicated by the fine black line. The white section of the plot corresponds to the nominal 2.5 GHz bandwidth of the XFFTS. The response of the lines in the second Nyquist band are shown in the grey zones. The red points are for the spectrum with the methanol at 50 μbar. The black points are for the baseline reference spectrum taken with an empty cell.

The signal is weak close to $2f_{LO}$ due to cut-off of the amplifier and increases quickly to provide a dynamic range in the order of 12 dB over most of the band, with a peak of 14 dB around 1.4 GHz. A small increase is observed at the start of the second Nyquist band, the response falls above 2.7 GHz. The data treatment is limited to 85 modes to ensure a minimum dynamic range of around 5 dB. The regularly spaced points situated approximately 3 dB above this main curve of fig. 4 correspond to values of $f_{rep}$ where two modes are at the same IF. The frequency attributed to each data point is $nf_{rep}$, so is not dependent on either $f_{LO}$ or $f_{IF}$.

Unlike conventional heterodyne detection with sources like black body radiation, the frequency accuracy of the THz-FC spectrometer is then totally independent of the frequency accuracy of the LO, and the use of a THz source whose frequency may not be precisely known like new molecular lasers should be possible[20]. Nevertheless, the stability of $f_{LO}$ will contribute to the overall signal to noise ratio and repeatability of the measured spectra. The transmission spectrum is directly calculated by normalisation of the measured spectrum by the baseline. A spectrum with continuous and uniform coverage is obtained. Here with methanol a large number of narrow lines are observed, fig. 5.

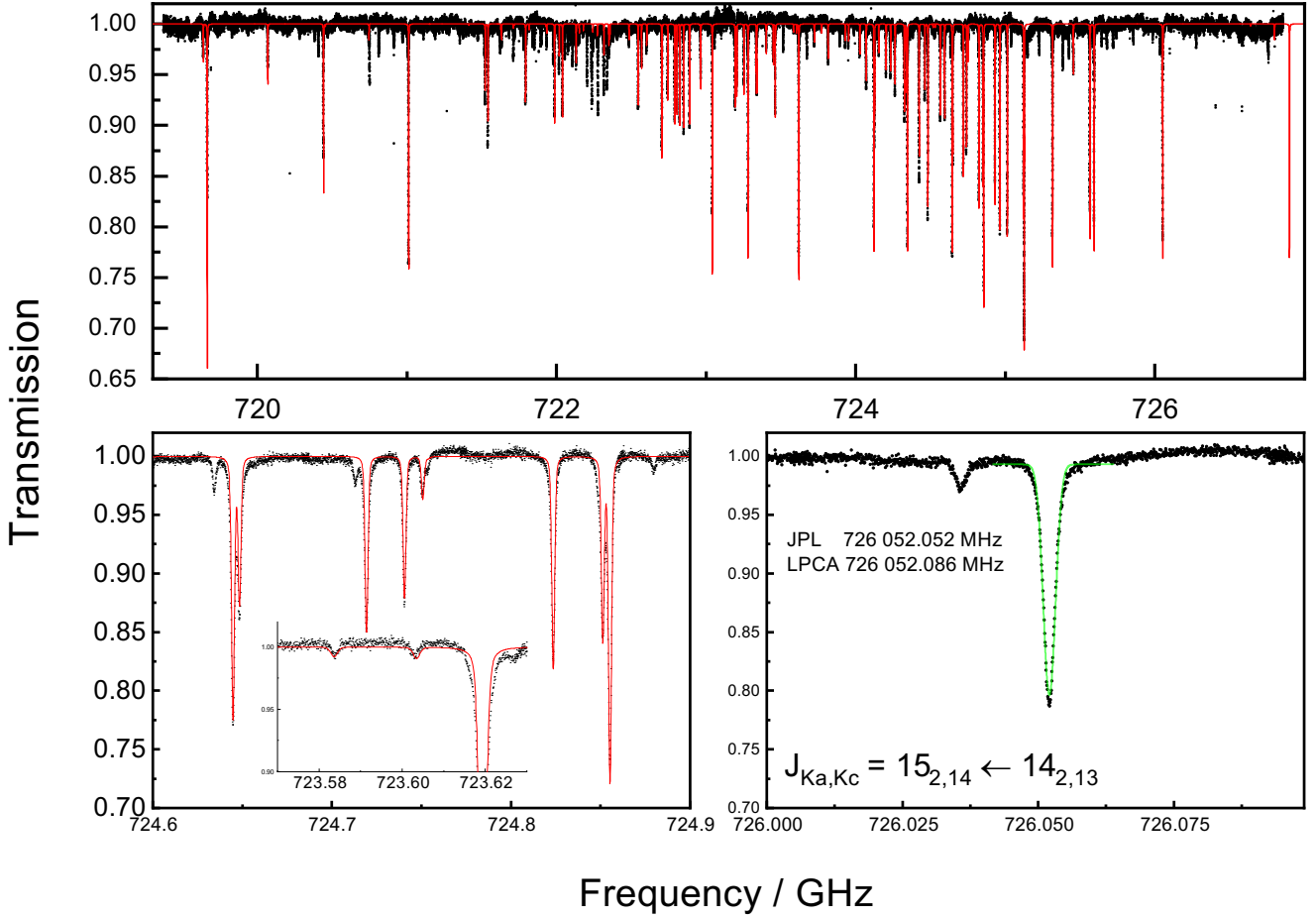

Fig 5. Transmission spectrum of methanol at a pressure of 50 mbar. Black points: measured data. Red line: approximation of a Voigt absorption profile calculated from the parameters tabulated in the HITRAN database. Green line: fitted Voigt profile for the methanol line at 726.052 GHz ( $J_{Ka,Kc} = 15_{2,14} \leftarrow 14_{2,13}$ (E species) $\nu_t = 0$ ).

In excess of 150 molecular transitions of methanol are observed in this spectrum, and some 80 of these lines can be immediately identified by the comparison with the HITRAN and JPL databases [21], [18]. Indeed, the methanol catalogue obtained from the study *by L. H. Xu et al.* [22] is limited in torsional states to $\nu_t \leq 2$ and rotational limits of $J \leq 35$ and $K \geq 15$ to avoid any extrapolation. Two more detailed zones are shown in the lower panels. The lower left-hand panel shows two doublets the first at 724.65 GHz and the second at 724.85 GHz, in both cases these close lines, separated by 3 MHz, are fully resolved. Two weak absorptions at 723.603 GHz and 723.584 GHz are clearly distinguished and are shown in the inset. Here the absorption is around 1%, and is close to the detection limit for this configuration. These transitions have lines strength in the order of $1 \times 10^{-23}$ cm$^{-1}$/(molecule.cm$^{-2}$). In the lower right hand panel a line is fitted with a Voigt profile to extract the measured line center, the frequency obtained is 30 kHz from the value tabulated in the JPL database [18]. This result can be explained by the precision of the model used to calculate the predicted lines in the catalogue, where the experimental rotational lines were fitted within 2 to 3 times their estimated experimental accuracy (ranging from 50 kHz to 200 kHz) [22].

IV. CONCLUSION

The combination of a THz-FC and a high frequency electronic mixer has enabled a new solution to the spectral resolution versus coverage compromise that faces designers of THz spectrometers. The simultaneous measurement of multiple THz-FC modes is a realistic solution. In its present configuration the resolution and measurement bandwidth are limited by the XFFTS, both of these can be overcome. The useable spectral resolution of 76 kHz is based on the standard channel spacing of the XFFTS, a finer resolution may be obtained by reconfiguring the channel spacing, this will however also reduce the bandwidth and increase the acquisition time required to achieve an identical signal to noise ratio. The base band of the XFFTS covering frequencies from 0 to 2.5 GHz provides a coverage of 5 GHz. We have shown this can be extended to by using the second Nyquist band of the XFFTS. Further extension is possible by using multiple XFFTS acquisition cards together. This would enable a larger bandwidth to be obtained with no time penalty. The IF bandwidth will ultimately be limited by the heterodyne mixer which can be above 40 GHz. For example, the next generation of FFTS is under development (qFFTS4G) and promises 4 inputs each with 4 GHz of instantaneous bandwidth, it can be used to cover 16 GHz of IF bandwidth (4x64k ≈ 256 kchannels) corresponding to 32 GHz in the THz band with a frequency resolution of 61 kHz [23]. This approach could increase the speed of high-resolution molecular spectrometers by an unprecedented factor paving the way for fundamental studies, or complex analyses like identification and quantification of a mixture of molecules.

REFERENCES

[1] F. Simon et al., 'Unrivaled accuracy in measuring rotational transitions of greenhouse gases: THz CRDS of $CF_4$', *Phys. Chem. Chem. Phys.*, p. 10.1039.D4CP00653D, 2024, doi: 10.1039/D4CP00653D.

[2] C. F. Neese, I. R. Medvedev, G. M. Plummer, A. J. Frank, C. D. Ball, and F. C. D. Lucia, 'Compact Submillimeter/Terahertz Gas Sensor With Efficient Gas Collection, Preconcentration, and ppt Sensitivity', *IEEE Sensors Journal*, vol. 12, no. 8, pp. 2565–2574, Aug. 2012, doi: 10.1109/JSEN.2012.2195487.

[3] F. Hindle et al., 'Monitoring of food spoilage by high resolution THz analysis', *Analyst*, vol. 143, no. 22, pp. 5536–5544, Nov. 2018, doi: 10.1039/C8AN01180J.

[4] N. Rothbart, O. Holz, R. Koczulla, K. Schmalz, and H.-W. Hübers, 'Analysis of Human Breath by Millimeter-Wave/Terahertz Spectroscopy', *Sensors*, vol. 19, no. 12, Art. no. 12, Jan. 2019, doi: 10.3390/s19122719.

[5] B. J. Drouin, F. W. Maiwald, and J. C. Pearson, 'Application of cascaded frequency multiplication to molecular spectroscopy', *Review of Scientific Instruments*, vol. 76, no. 9, p. 093113, Sep. 2005, doi: 10.1063/1.2042687.

[6] F. Hindle, R. Bocquet, A. Pienkina, A. Cuisset, and G. Mouret, 'Terahertz gas phase spectroscopy using a high-finesse Fabry–Pérot cavity', *Optica*, vol. 6, no. 12, p. 1449, Dec. 2019, doi: 10.1364/OPTICA.6.001449.

[7] E. R. Brown, K. A. McIntosh, K. B. Nichols, and C. L. Dennis, 'Photomixing up to 3.8 THz in low-temperature-grown GaAs', *Applied Physics Letters*, p. 285, 1995, doi: 10.1063/1.113519.

[8] W. Nsengiyumva et al., 'Sensing and Nondestructive Testing Applications of Terahertz Spectroscopy and Imaging Systems: State-of-the-Art and State-of-the-Practice', *IEEE Transactions on Instrumentation and Measurement*, vol. 72, pp. 1–83, 2023, doi: 10.1109/TIM.2023.3318676.

[9] S. Eliet, A. Cuisset, F. Hindle, J.-F. Lampin, and R. Peretti, 'Broadband Super-Resolution Terahertz Time-Domain Spectroscopy Applied to Gas Analysis', *IEEE Trans. THz Sci. Technol.*, vol. 12, no. 1, pp. 75–80, Jan. 2022, doi: 10.1109/TTHZ.2021.3120029.

[10] S. A. Diddams et al., 'Direct RF to optical frequency measurements with a femtosecond laser comb', *IEEE Transactions on Instrumentation and Measurement*, vol. 50, no. 2, pp. 552–555, Apr. 2001, doi: 10.1109/19.918189.

[11] H. Schnatz et al., 'Optical frequency measurements using fs-comb generators', *IEEE Transactions on Instrumentation and Measurement*, vol. 54, no. 2, pp. 750–753, Apr. 2005, doi: 10.1109/TIM.2004.843417.

[12] T. M. Goyette et al., 'Femtosecond demodulation source for high-resolution submillimeter spectroscopy', *Applied Physics Letters*, vol. 67, no. 25, pp. 3810–3812, Dec. 1995, doi: 10.1063/1.115391.

[13] Y.-D. Hsieh et al., 'Spectrally interleaved, comb-mode-resolved spectroscopy using swept dual terahertz combs', *Sci Rep*, vol. 4, no. 1, p. 3816, Jan. 2014, doi: 10.1038/srep03816.

[14] T. Yasui et al., 'Super-resolution discrete Fourier transform spectroscopy beyond time-window size limitation using precisely periodic pulsed radiation', *Optica, OPTICA*, vol. 2, no. 5, pp. 460–467, May 2015, doi: 10.1364/OPTICA.2.000460.

[15] J. L. Hall and J. Ye, 'Optical frequency standards and measurement', *IEEE Transactions on Instrumentation and Measurement*, vol. 52, no. 2, pp. 227–231, Apr. 2003, doi: 10.1109/TIM.2003.810450.

[16] A. S. Skryl, D. G. Pavelyev, M. Y. Tretyakov, and M. I. Bakunov, 'High-resolution terahertz spectroscopy with a single tunable frequency comb', *Opt. Express, OE*, vol. 22, no. 26, pp. 32276–32281, Dec. 2014, doi: 10.1364/OE.22.032276.

[17] J. K. Messer, F. C. De Lucia, and P. Helminger, 'Submillimeter spectroscopy of the major isotopes of water', *Journal of Molecular Spectroscopy*, vol. 105, no. 1, pp. 139–155, May 1984, doi: 10.1016/0022-2852(84)90109-7.

[18] H. M. PICKETT, R. L. POYNTER, E. A. COHEN, M. L. DELITSKY, J. C. PEARSON, and H. S. P. MÜLLER, 'SUBMILLIMETER, MILLIMETER, AND MICROWAVE SPECTRAL LINE CATALOG', *Journal of Quantitative Spectroscopy and Radiative Transfer*, vol. 60, no. 5, pp. 883–890, Nov. 1998, doi: 10.1016/S0022-4073(98)00091-0.

[19] B. Klein, S. Hochgürtel, I. Krämer, A. Bell, K. Meyer, and R. Güsten, 'High-resolution wide-band fast Fourier transform spectrometers', *A&A*, vol. 542, p. L3, Jun. 2012, doi: 10.1051/0004-6361/201218864.

[20] M.-H. Mammez et al., 'Optically Pumped Terahertz Molecular Laser: Gain Factor and Validation up to 5.5 THz', *Advanced Photonics Research*, p. 2100263, Jan. 2022, doi: 10.1002/adpr.202100263.

[21] I. E. Gordon et al., 'The HITRAN2020 molecular spectroscopic database', *Journal of Quantitative Spectroscopy and Radiative Transfer*, vol. 277, 2022, doi: 10.1016/j.jqsrt.2021.107949.

[22] L.-H. Xu et al., 'Torsion–rotation global analysis of the first three torsional states ($v_t$ = 0, 1, 2) and terahertz database for methanol', *Journal of Molecular Spectroscopy*, vol. 251, no. 1, pp. 305–313, Sep. 2008, doi: 10.1016/j.jms.2008.03.017.

[23] B. Klein, S. Hochgürtel, I. Krämer, A. Bell, and G. Grutzeck, 'Digital high-resolution wide-band fast Fourier transform spectrometer', in *Proceedings of ISSTT 2019 - 30th International Symposium on Space Terahertz Technology*, Gothenburg, 2019, p. 226.